# The Magnus (Kutta-Jukovskii) Force Acting on a Sphere

L. Kiknadze,  Yu. Mamaladze

E. Andronikashvili Institute of Physics, Tbilisi, Georgia
Tamarashvili 6, Tbilisi, 0177 Georgia
yum@iphac.ge

Fluid Dynamics

The expression is derived for the Magnus force acting on the sphere.

The Magnus force (known also as Kutta-Jukovskii lifting force) acts on a cylindrical body if ideal liquid flows across its axis and, simultaneously, around it (see below Eq.(2)). This force does not depend on the shape of the body cross-section and, being reduced to the unit length, is written down as (see for example Lamb 1945):

$$\vec{F}_1 = \rho \vec{\Gamma} \times (\vec{v}_0 - \vec{v}). \tag{1}$$

Here $\rho$ is the density of the liquid, $\vec{\Gamma}$ is the circulation, $\vec{v}_0$ and $\vec{v}$ are the velocities of the body and the flow, respectively. We could not find the corresponding expression for a spherical body in literature. Maybe, this problem has not been put forward, because the circulation cannot be located – it must be continued beyond the sphere. A vortex line, containing a sphere, is a very unnatural object to be considered in the hydrodynamics of the classical ideal liquid. But, it is quite usual for the quantum hydrodynamics of the superfluid HeII, where vortex lines, with an ion in their cores are well studied (Raifeld and Raif 1964, Ashton and Glaberson 1979, Donnelly 1991).

Keeping in mind the existence of such objects, let us consider the motionless sphere with the radius $R$, and the flow of the ideal incompressible liquid along $y$ - axis with the velocity $u$ and around the $z$ - axis with the circulation $\Gamma$ (the origin of coordinates is at the center of the sphere). The potential of flow velocity is:

$$\varphi = \frac{R^3}{2r^3}uy + uy + \frac{\Gamma}{2\pi} arctg \frac{y}{x} \tag{2}$$

where $r = (x^2 + y^2 + z^2)^{1/2}$, and the force acting on the sphere surface $S$ is

$$\vec{F} = -\int_{(S)} P\vec{v}dS. \tag{3}$$

Here $\vec{v} = \vec{R}/R$ and

$$P = P_0 - \frac{1}{2}\rho(\vec{\nabla}\varphi)^2. \tag{4}$$

It is convenient to perform the integration in Eq. (3) using the spherical coordinates $r, \theta, \alpha$ $(x = r\sin\theta\cos\alpha, y = r\sin\theta\sin\alpha, z = r\cos\theta)$. This implies:

$$F_x = \frac{3}{4\pi}\rho\Gamma uR \int_0^{2\pi} \cos^2\alpha\, d\alpha \int_0^\pi \sin\theta\, d\theta = \frac{3}{2}\rho\Gamma uR, \tag{5}$$

$$F_y = \frac{3}{4\pi}\rho\Gamma uR \int_0^{2\pi} \cos\alpha \sin\alpha\, d\alpha \int_0^\pi \sin\theta\, d\theta = 0, \tag{6}$$

$$F_z = \frac{3}{4\pi}\rho\Gamma uR \int_0^{2\pi} \cos\alpha\, d\alpha \int_0^\pi \cos\theta\, d\theta = 0. \tag{7}$$

This result can be rewritten in the vector form:

$$\vec{F} = \frac{3}{2}\rho\vec{\Gamma} \times \left(\vec{v_0} - \vec{v}\right). \tag{8}$$

The absence of the Magnus force in the cases of free motion where a body (it may also be a vortex) transfers together with the media with its (the media's) velocity: $\vec{v_0} = \vec{v}$ is ensured by the common factor ($\vec{v_0} - \vec{v}$) in Eqs. (1) and (8). In particular, the free vortex moves exactly in this way. For example, in the above- mentioned objects - the complex "vortex + ion" - the each element of a vortex moves with the velocity that superfluid has at the place of this element's location except the rotation around it ($\vec{v_0} = \vec{v}$). The captured ion is transferred together with the vortex and the Magnus forces acting on the vortex and on the ion are zero. The equality $\vec{v_0} = \vec{v}$ is broken due to the action of electric or friction forces. In such cases the Magnus force is not zero.